\documentclass{article}

\PassOptionsToPackage{numbers, compress}{natbib}

\usepackage[preprint]{neurips_2026}
\makeatletter
\renewcommand{\@noticestring}{Accepted at the 40th Conference on Neural Information Processing Systems (NeurIPS 2026).}
\makeatother
\usepackage[utf8]{inputenc}
\usepackage[T1]{fontenc}
\usepackage{hyperref}
\usepackage{url}
\usepackage{booktabs}
\usepackage{amsfonts}
\usepackage{amsmath}
\usepackage{amssymb}
\usepackage{nicefrac}

\newenvironment{acknowledgements}{\section*{Acknowledgements}}{}
\usepackage{microtype}
\usepackage{xcolor}
\usepackage{graphicx}
\usepackage{algorithm}
\usepackage{algorithmic}
\usepackage{multirow}
\usepackage{subcaption}
\usepackage{pgfplots}
\pgfplotsset{compat=1.18}
\usepackage{multirow}
\usepackage{makecell}
\usepackage[capitalize,sort,compress]{cleveref}

\usepackage{safecolours}
\usepackage{xspace}

\usetikzlibrary{decorations.pathreplacing}
\definecolor{todo}{HTML}{D55E00}

\newcommand{\lse}{\operatorname{LSE}}
\definecolor{fuselseColor}{HTML}{EE6677}  
\definecolor{intlseColor}{HTML}{4477AA}   
\definecolor{scTeal}{HTML}{44AA99}
\newcommand{\fuselse}{\textcolor{fuselseColor}{\textsc{\bf FuseLSE}}\xspace}
\newcommand{\intlse}{\textcolor{intlseColor}{\textsc{\bf IntLSE}}\xspace}

\newcommand{\ie}{\textit{i.e.}\xspace}

\title{Speeding up Log-Sum-Exp: Kernel Fusion at the Memory Wall,
Integer Arithmetic at the Compute Wall}

\author{%
   Lingyun Yao \thanks{Correspondence to: \texttt{lingyun.yao@aalto.fi}. Code: \url{https://github.com/lingyunyao/fuselse}}\\
  Aalto University\\
  Espoo, Finland  \\
  \And
     Martin Andraud\\
  UC Louvain\\
  Louvain-la-Neuve, Belgium   \\
  \And
    Niki Loppi \\
  NVIDIA\\
  Helsinki, Finland  \\
  \And
    Andrea Pilzer \\
  NVIDIA\\
  Bologna, Italy   \\
  \And
     Anji Liu \\
  National University of Singapore \\
  Singapore    \\
  \And
   Guy Van den Broeck \\
  University of California\\
  Los Angeles, USA  \\
  \And
    Martin Trapp \\
  KTH Royal Institute of Technology\\
  Stockholm, Sweden   \\
}

\begin{document}
\maketitle

\begin{abstract}
While the Log-Sum-Exp (LSE) function is a foundational numerical primitive for stable log-domain computation in modern machine learning, its fast and efficient computation on generic platforms (\ie, CPUs and GPUs) has been underexplored, potentially creating execution bottlenecks for large workloads. On GPU, the de facto implementation \texttt{torch.logsumexp} dispatches a chain of nine sub-kernels per call, incurring redundant High Bandwidth Memory (HBM) traffic and launch overhead. On CPU, \texttt{exp} and \texttt{log} operations are hardware-hungry, limiting the computing efficiency. To address these bottlenecks, we propose two kernels to significantly accelerate LSE computation on both GPUs and CPUs: \fuselse{}, a single-pass fused kernel that evaluates LSE's \texttt{exp} and \texttt{log} functions on the GPU's Special Function Unit (SFU); and \intlse{}, which further replaces the LSE computation with approximated integer arithmetic for CPU computation. Our experiments, validated by benchmarks across workload sizes and hardware platforms, show that both kernels are ${\sim}10\!-\!11\times$ faster than PyTorch at small workloads and ${\sim}6\times$ faster on larger ones on GPUs, where the speed-up is attributed to reduced launch overhead and HBM traffic. On CPU, where no SFU is available, LSE computation
dominates the cost, allowing \intlse{} to gain ${\sim}3\!-\!6\times$ execution speed over \fuselse{}. 
\end{abstract}

\section{Introduction}
\label{sec:intro}

The Log-Sum-Exp (LSE) function is one of the most widespread numerical primitives in modern machine learning, whenever numerically stable logarithmic computation is required. For example, LSE is used to compute per-token log-probabilities of Large Language Models (LLMs) during reinforcement learning, where every token's forward pass calls one LSE over the vocabulary size~\cite{vonwerra2020trl}; to normalize particle weights at every time step of a particle filter computation~\cite{andrieu2010particle, johansen2009tutorial, chopin2020introduction}; to enable log-domain probabilistic inference, for instance in probabilistic circuits~\cite{peharz2020einsum, choi2020probabilistic, liu2024scaling}; or to stabilize the Sinkhorn algorithm for entropy-regularized optimal transport~\cite{gabriel2019computational, schmitzer2019stabilized}. 

Despite this ubiquity, LSE itself is rarely an optimization target on generic CPUs and GPUs, where efforts focused on dense matrix multiplications and standard reduction primitives. To illustrate this gap, PyTorch's \texttt{torch.logsumexp} function can be $\sim\!6\times$ slower than the analogous \texttt{torch.sum} reduction over the same tensor on a given GPU ($1.95$\,ms vs.\ $0.31$\,ms in \cref{fig:torch-profile}). On GPUs, although log computation libraries exist, they either support only element-wise operations~\cite{xlnscuda} based on Mitchell's approximation~\cite{mitchell}, or are not fully optimized for LSE~\cite{torch-compile,tillet2019triton,milakov2018online,dao2022flashattention}. On CPUs, ML applications requiring LSE rely on generic libraries (\texttt{scipy},\texttt{numpy}, \texttt{torch}) (\cref{sec:exp-pf}), all of which implement LSE via the same multi-step primitives. However, LSE optimization has shown promising gains to push the efficiency of specialized hardware for AI inference, including efficient FPGA implementations~\cite{weber2019resource}, integer approximations on 16\,nm CMOS~\cite{yao2025logsumexp,yao2026einchip}, or alternative arithmetic in place of log-domain math on dedicated hardware~\cite{shah2021dpu,pmlr-v244-yao24a, periasamy2024autopc}.

Considering the growing model sizes and interest around probabilistic and symbolic models that perform extensive log-space computation~\cite{garcez2023neurosymbolic}, LSE plays an ever-increasing role in optimizing their training and inference time. Hence, this work targets the optimization of LSE on both CPU and GPU by exploiting two key opportunities. First, current baseline implementations dispatch a chain of sub-kernels per call, each paying a fixed launch overhead. Second, four of these sub-kernels touch the full input tensor and round-trip it through High-Bandwidth Memory (HBM) roughly $6\times$ per call (\cref{sec:profile}). We instantiate two acceleration strategies as kernel replacements: \fuselse{}, which fuses the sub-kernel chain to eliminate the launch overhead and reduce HBM traffic, and \intlse{}, which further replaces the \texttt{exp} and \texttt{log} functions with integer approximations~\cite{yao2025logsumexp, mitchell}. 

In terms of results, our experiments show that both kernels achieve a ${\sim}10\!-\!11\times$ speed-up over \texttt{torch.logsumexp} at small workloads and ${\sim}6\times$ at large workloads, confirmed by single-kernel benchmarks and three ML applications. We trace this gain through a time-breakdown analysis across workload sizes, GPUs with varying HBM bandwidths, and input distributions. Our results show that the choice of arithmetic is a second-order factor on GPU: \fuselse{} and \intlse{} achieve nearly identical speed, and both outpace HBM transfers. On CPU, where no Special Function Unit (SFU) is available, the choice can no longer be ignored: \intlse{}'s integer pipeline beats \fuselse{} by $3$--$6\times$ on the microbenchmark and \texttt{torch.logsumexp} by $2.8$--$5.2\times$ for particle filters.

\section{Preliminaries}
\label{sec:prelim}
The LSE function is defined as:
\begin{equation}
\label{eq:lse-def}
\lse(x_1, \ldots, x_N) \;\triangleq\; c + \log \Bigl(\textstyle\sum_{i=1}^{N} e^{x_i - c}\Bigr), \qquad \text{with} \quad c = \max_i\, x_i  .
\end{equation} 
LSE arises whenever probabilities or scores must be aggregated in the logarithmic domain for numerical stability. We highlight three application domains where LSE is a potential computational bottleneck, each evaluated experimentally in \cref{sec:exp-rlhf,sec:exp-pyjuice,sec:exp-pf}.

\textbf{RLHF log-probability computation.}\;
Reinforcement learning from human feedback (RLHF) \cite{lambert2025reinforcement} methods such as PPO \cite{schulman2017proximal}, and
RLOO \cite{ahmadian2024back} require per-token log-probabilities $\log p(y_t \mid y_{<t}, x) = z_{y_t} - \lse(z_1, \dots, z_V)$ at every training step for both the policy and reference models.
The vocabulary dimension $V$ ranges from $32k$ (Llama-2) to $256k$ (Gemma), making the $\lse$ reduction over $V$ expensive.
The HuggingFace TRL library \cite{vonwerra2020trl} implements this via \texttt{selective\_log\_softmax}, in which \texttt{torch.logsumexp} accounts for a significant share of runtime.

\textbf{Probabilistic circuits.}\;
Probabilistic circuits (PCs) \cite{choi2020probabilistic} are a family of tractable probabilistic models that represent probability distributions as directed
acyclic computation graphs with sum ($\oplus$) and product ($\otimes$) nodes.
When computed in the log domain to avoid numerical underflow, product nodes become additions and sum nodes require LSE operations \cite{peharz2020einsum}.
A single inference pass through a deep PC may invoke millions of LSE evaluations \cite{liu2024scaling}. 
Here we use PyJuice~\cite{liu2024scaling}, a GPU-accelerated library for PC inference and learning, as both inference and learning heavily rely on LSE.

\textbf{Particle filtering.}\;
Particle filters \cite{johansen2009tutorial,chopin2020introduction} maintain a weighted set of particles to approximate the posterior, for example, of Bayesian state-space models.
Standard implementations require computing normalized log weights, which are obtained via $\log \tilde{w}_i = \log w_i - \lse(\log w_1, \dots, \log w_N)$, making one LSE reduction per time step a fixed cost of the filter loop.
In this work, we focus on the bootstrap particle filter \cite{gordon1993novel} (BPF), a widely used SMC instance, implemented in the particles library \cite{code_particles}.

\subsection{Baseline: PyTorch's \texttt{torch.logsumexp}}
\label{sec:profile}
PyTorch's LSE implementation~\cite{pytorch-logsumexp-docs} follows the formulation of \cref{eq:lse-def}, \ie, 
(i)~compute $c = \max(x)$;
(ii)~subtract the max, $x - c$; 
(iii)~exponentiate;
(iv)~sum along the reduction dimension; 
(v)~take the log;
(vi)~add the max back.
To check how this composes on the GPU, we trace a single call of \texttt{torch.logsumexp(x, dim=-1)} at $(B,N)\!=\!(512,\,128256)$ on a V100 with \texttt{torch.profiler} (20 timed iterations, 50-step warm-up).
The six-step recipe expands into a chain of nine CUDA sub-kernels shown in \cref{fig:torch-profile}.
Extra \texttt{abs}, \texttt{unary}, and \texttt{mask\_fill} kernels handle the $\pm\infty$ corner cases of step~(i) and are only visible as thin lines in the profile due to their negligible contributions.
Nsight Compute shows that the following four sub-kernels dominate the runtime (\texttt{amax}, \texttt{sub}, \texttt{exp}, and \texttt{sum}, at
$312$, $655$, $642$, and $317\,\mu$s respectively), and these same four account for six $(B,N)$-sized HBM transfers: \texttt{amax} $1\times$ (read), \texttt{sub} $2\times$ (read+write), \texttt{exp} $2\times$ (read+write), and \texttt{sum} $1\times$ (read). 

\subsection{Existing accelerating methods} 
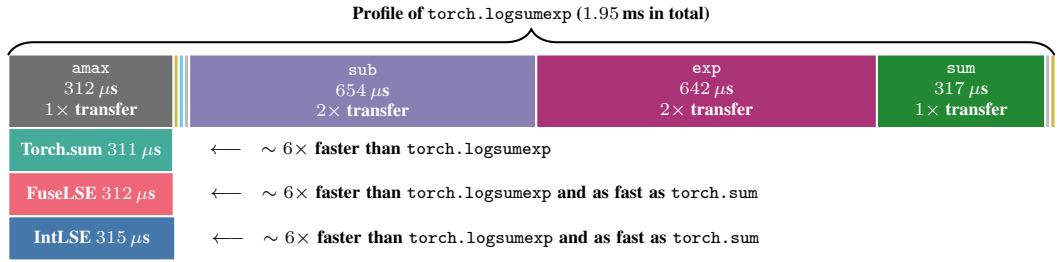
\begin{figure}[!htbp]
  \centering
\noindent\resizebox{\linewidth}{!}{%
\centering
\begin{tikzpicture}[
    minimum height=1cm, anchor=west, draw=white, thick, inner sep=0pt, font=\bfseries\scriptsize, color=white
]
    \newlength{\segwidth}
    \newlength{\xoffset}

    \newcommand{\calcwidth}[1]{%
      \pgfmathsetlength{\segwidth}{#1/1937.14*\linewidth}%
      \ifdim\segwidth<2pt \setlength{\segwidth}{2pt}\fi
    }

    \setlength{\xoffset}{0pt}
    \calcwidth{311.51}%
    \node[fill=scGrey!60!black, draw=white, thick, 
        minimum width=\segwidth, align=center] at (\xoffset,0) {\texttt{amax}\\$312\,\mu$s\\$1\times$ transfer};%
    \addtolength{\xoffset}{\segwidth}%

    \calcwidth{2.35}%
    \node[fill=scYellow, draw=white, thick, 
        minimum width=\segwidth, align=center] at (\xoffset,0) {};%
    \addtolength{\xoffset}{\segwidth}%

    \calcwidth{2.17}%
    \node[fill=scCyan, draw=white, thick, 
        minimum width=\segwidth, align=center] at (\xoffset,0) {};%
    \addtolength{\xoffset}{\segwidth}%

    \calcwidth{2.04}%
    \node[fill=scGrey, draw=white, thick, 
        minimum width=\segwidth, align=center] at (\xoffset,0) {};%
    \addtolength{\xoffset}{\segwidth}%

    \calcwidth{654.30}%
    \node[fill=scCyan!50!scPurple, draw=white, thick, 
        minimum width=\segwidth, align=center] at (\xoffset,0) {\texttt{sub}\\$654\,\mu$s\\$2\times$ transfer};%
    \addtolength{\xoffset}{\segwidth}%

    \calcwidth{642.18}%
    \node[fill=scPurple, draw=white, thick, 
        minimum width=\segwidth, align=center] at (\xoffset,0) {\texttt{exp}\\$642\,\mu$s\\$2\times$ transfer};%
    \addtolength{\xoffset}{\segwidth}%

    \calcwidth{317.20}%
    \node[fill=scGreen, draw=white, thick, 
        minimum width=\segwidth, align=center] at (\xoffset,0) {\texttt{sum}\\$317\,\mu$s\\$1\times$ transfer};%
    \addtolength{\xoffset}{\segwidth}%

    \calcwidth{3.12}%
    \node[fill=scGrey, draw=white, thick, 
        minimum width=\segwidth, align=center] at (\xoffset,0) {};%
    \addtolength{\xoffset}{\segwidth}%

    \calcwidth{2.27}%
    \node[fill=scYellow, draw=white, thick, 
        minimum width=\segwidth, align=center] at (\xoffset,0) {};%

    \draw [decorate, black, decoration={brace, amplitude=10pt, raise=4pt}]
    (0.01,0.35) -- (\xoffset+\segwidth,0.35) node [midway, above=5pt] {Profile of \texttt{torch.logsumexp} ($1.95$\,ms in total)};

    \begin{scope}[yshift=-0.8cm, minimum height=0.6cm]
        
        \setlength{\xoffset}{0pt}
        \calcwidth{311}%
        \node[fill=scTeal, draw=white, thick, 
            minimum width=\segwidth, align=center] at (\xoffset,0) {Torch.sum $311\,\mu$s};%
        \addtolength{\xoffset}{\segwidth}%
        \calcwidth{1625}%
        \node[fill=white, draw=white, thick, 
            minimum width=\segwidth, text width=\segwidth, align=left] at (\xoffset,0) {\textcolor{black}{\qquad$\longleftarrow \quad\sim 6\times$ faster than \texttt{torch.logsumexp}}
            };%
    \end{scope}

    \begin{scope}[yshift=-1.4cm, minimum height=0.6cm]
        
        \setlength{\xoffset}{0pt}
        \calcwidth{312}%
        \node[fill=scRed, draw=white, thick, 
            minimum width=\segwidth, align=center] at (\xoffset,0) {FuseLSE $312\,\mu$s};%
        \addtolength{\xoffset}{\segwidth}%
    
        \calcwidth{1625}%
        \node[fill=white, draw=white, thick, 
            minimum width=\segwidth, text width=\segwidth, align=left] at (\xoffset,0) {\textcolor{black}{\qquad$\longleftarrow \quad\sim 6\times$ faster than \texttt{torch.logsumexp} and as fast as \texttt{torch.sum}}
            };%
        
    \end{scope}
    \begin{scope}[yshift=-2cm, minimum height=0.6cm]
        
        \setlength{\xoffset}{0pt}
        \calcwidth{315}%
        \node[fill=scBlue, draw=white, thick, 
            minimum width=\segwidth, align=center] at (\xoffset,0) {IntLSE $315\,\mu$s};%
        \addtolength{\xoffset}{\segwidth}%
    
        \calcwidth{1622}%
        \node[fill=white, draw=white, thick, 
            minimum width=\segwidth, text width=\segwidth, align=left] at (\xoffset  ,0) {\textcolor{black}{\qquad$\longleftarrow \quad\sim 6\times$ faster than \texttt{torch.logsumexp} and as fast as \texttt{torch.sum}}
            };%
        
    \end{scope}

\end{tikzpicture}
}
  \caption{%
    Per-call profile at $(B,N)\!=\!(512,128256)$, on V100-PCIe-32\,GB. Top
    row: \texttt{torch.logsumexp} with nine CUDA
    sub-kernels; the four heavy ops (\texttt{amax}, \texttt{sub},
    \texttt{exp}, \texttt{sum}) are labeled with kernel name, GPU
    time, and HBM traffic as a multiple of the input volume, and the five narrow ops
    (\texttt{abs}, \texttt{unary}, \texttt{mask\_fill}, \texttt{log},
    \texttt{add}) are shown as thin coloured stripes.
    Bottom three rows: \texttt{torch.sum} for reference, \fuselse{}, and \intlse{} using only one kernel. Bar widths are linear in time. Per-launch times from \texttt{torch.profiler}; HBM traffic data from Nsight Compute.}
  \label{fig:torch-profile}
\end{figure}

The \texttt{torch.compile} operation (TorchInductor~\cite{torch-compile}) fuses the \texttt{torch.logsumexp} operation chain at runtime into Triton~\cite{tillet2019triton} kernels. On the workload of \cref{fig:torch-profile}, \texttt{torch.compile(torch.logsumexp)} delivers a $\sim 3\times$ speedup and cuts HBM traffic from $\sim 6\times$ to $\sim 2\times$ the input size by eliminating inter-operator HBM traffic. However, it inherits PyTorch's two-pass max-subtract semantics and pays recompilation cost on every new input shape. TensorFlow's \texttt{tf.math.reduce\_logsumexp}~\cite{tensorflow-logsumexp-docs} and JAX's \texttt{jax.nn.logsumexp}~\cite{jax-logsumexp-docs} define the same two-pass structure, which the XLA compiler~\cite{openxla} preserves when compiling them.
Online softmax~\cite{milakov2018online} computes LSE in a single streaming pass by carrying two scalars (max and exp-sum); FlashAttention~\cite{dao2022flashattention} exploits this to fuse softmax with the matmuls of self-attention, CCE~\cite{wijmans2024cut} fuses LSE with the LM-head matmul, PyJuice~\cite{liu2024scaling} performs the same fusion for dense PCs. We benchmarked against this matmul-LSE-fused method in \cref{sec:exp-rlhf}, the LM-head setting where they apply. PyJuice's sparse mode~\cite{liu2024scaling}, EinsumNetworks (einsum-based PC framework)~\cite{peharz2020einsum}, and particle filters~\cite{code_particles} fall back to the two-pass max-subtract recipe in the absence of an upstream matmul; these settings require an optimized standalone LSE primitive.
\vspace{-1.5em}
\vspace{-1em}
\section{Implementation of the proposed LSE acceleration kernels}
\label{sec:method}
The profile in \cref{fig:torch-profile} suggests two design opportunities: (1) fusing the nine kernel launches into one removes eight launches and reduces HBM traffic from $\sim 6\times$ to $\sim 1\times$ the input size, and (2) replacing \texttt{exp} and \texttt{log} by cheaper approximate operations, e.g., integer arithmetic~\cite{yao2025logsumexp}. These motivate our two kernels, \fuselse{} (\cref{sec:fuselse}) and \intlse{} (\cref{sec:intlse}), whose details are outlined below. 
\subsection{FuseLSE: Single-Pass Fused Kernel}
\label{sec:fuselse}
We base \fuselse{} on the pair-wise form of log-sum-exp. Setting $N=2$ in \cref{eq:lse-def} with inputs $x, y$ and $c=\max(x,y)$ gives
\begin{equation}
    \lse(x, y) = \max(x, y) + \log\!\bigl(1 + \exp(-|x{-}y|)\bigr), \label{eq:lse-pair-float}
\end{equation}
since the two exponents become $0$ and $-|x-y|$. As $\lse$ is associative and commutative, applying \cref{eq:lse-pair-float} recursively in any order recovers the $N$-ary LSE.
Note that each evaluation carries its own local max operation, with no upfront pass over the data to compute a global max. 
In exchange, \fuselse{} evaluates $N{-}1$ \texttt{log} calls per row instead of one.

\fuselse{} takes a float32 matrix $X\!\in\!\mathbb{R}^{B\times N}$ of base-$e$ log-data ($B$ rows, each of length $N$) and returns the $B$ row-wise scalars in a single HBM pass. The kernel launches $B$ blocks, each with $256$ threads that cooperatively reduce one
$N$-element row to a single scalar. Each thread keeps one running accumulator in a register throughout, initialized to $-10^{30}$ and assigned the first valid input directly to avoid the NaN corner. Within a block, the $256$ threads sweep the $N$-element row through \texttt{float4} (4 of float32 data) loads: each load
delivers $4$ float32 per thread, and a warp of $32$ threads
covers $128$ float32 per step. With this, the kernel approaches peak HBM bandwidth. Each thread visits about $N/256$ floats and folds each into its
register-resident accumulator via Eq.~\eqref{eq:lse-pair-float}, evaluated through the SFU's of the GPU intrinsics \texttt{\_\_expf} and \texttt{\_\_logf}. 

The $256$ partials are reduced to one scalar via two stages of warp-shuffle butterflies: $32 \to 1$ within each warp, then $8 \to 1$ across warps through a $32$-byte shared-memory buffer. Thread~$0$ writes the final scalar to global memory. Overall, the kernel reads each input element once and writes $B$ output scalars, for a total HBM traffic of roughly $1\times$ the input size, against 6 times for \texttt{torch.logsumexp} (\cref{sec:profile}).

\subsection{IntLSE: Approximate integer LSE}
\label{sec:intlse}
\intlse{} preserves the structure of \fuselse{} and replaces only the pair function in \cref{eq:lse-pair-float} with integer arithmetic, following \citet{yao2025logsumexp}. Assuming $x \geq y$ and letting $z = y - x \leq 0$,
the pair function becomes $\lse(x, y) \;=\; x + \log_2\!\bigl(1 + 2^{z}\bigr)$.
Define $f(z) \triangleq \log_2(1 + 2^z)$. Applying Mitchell's approximations~\cite{mitchell},
$2^{v}\!\approx\!1{+}v$ and $\log_2(1{+}v)\!\approx\!v$ for
$v\!\in\![0, 1)$, $f(z)$ collapses to a single addition and a
right-shift,
\begin{equation}
\hat{f}(z) \;\triangleq\; (1 + F_z) \gg (-I_z),
\label{eq:combined-approx}
\end{equation}
where $I_z \le 0$ and $F_z \in [0, 1)$ are the integer and fractional parts of $z$. The pair function reduces to $\lse(x, y) \approx x + \hat{f}(z)$, computed with only integer adds and shifts. The error $\varepsilon(z) = f(z) - \hat{f}(z)$ is non-negative and periodic, peaking at each integer endpoint $z = -n$ with value $\log_2(1 + 2^{-n}) - 2^{-n}$ (\cref{fig:error-analysis} a,b); the global maximum $\approx 0.085$ occurs at $n = 1$.
Both $\hat{f}(z)$ and $\varepsilon(z)$ decay exponentially as $z \to -\infty$ (\cref{fig:error-analysis}c); distant elements contribute negligibly.  Assuming $z$ is sampled with unit density on $(-\infty, 0]$, the expected accumulated error is $\int_{-\infty}^{0}\!\varepsilon(z)\,dz \approx\; 0.212$, independent of sample size (\cref{app:error-details}), restricting the error
accumulated over a chain of LSE operations. The actual error accumulated over a row depends on the input distribution, and is measured empirically in \cref{tab:dist-accuracy}.
A $64$-entry constant-memory correction look-up table (LUT), stored in CUDA constant
memory, indexed by the top $6$ bits of $\hat{f}(z)$, further reduces the per-step peak error. The LUT lowers the maximum pointwise error from $0.085$ to $0.0068$ ( \cref{fig:error-analysis} b).

\definecolor{peak1}{RGB}{228,26,28}    
\definecolor{peak2}{RGB}{125,126,184}   
\definecolor{peak3}{RGB}{77,175,74}    
\definecolor{peak4}{RGB}{152,78,163}   
\definecolor{peak5}{RGB}{255,127,0}    
\begin{figure}[!htbp]
\centering
\begin{tikzpicture}[baseline=(current bounding box.north)]
\begin{axis}[
    width=0.24\textwidth,
    height=0.18\textwidth,
    scale only axis,
    xlabel={$z$},
    ylabel={Value},
    grid=both,
    grid style={line width=0.1pt, draw=gray!20},
    legend pos=north west,
    legend style={font=\tiny, fill=white, fill opacity=0.8, draw opacity=1, text opacity=1},
    xmin=-4, xmax=0,
    ymin=0, ymax=1.1,
    tick label style={font=\tiny},
    label style={font=\scriptsize},
    samples=200,
    domain=-4:0,
    title={(a) Exact vs. App.},
    title style={at={(0.5,1.02)}},
]
\addplot[color={rgb,255:red,0;green,119;blue,187}, thick, mark=none] {ln(1 + 2^x)/ln(2)};
\addlegendentry{$f(z)$}
\addplot[color={rgb,255:red,204;green,102;blue,0}, thick, mark=none] {(1 + x - floor(x)) * 2^(floor(x))};
\addlegendentry{$\hat{f}(z)$}
\end{axis}
\end{tikzpicture}%
\hfill%
\begin{tikzpicture}[baseline=(current bounding box.north)]
\begin{axis}[
    width=0.24\textwidth,
    height=0.18\textwidth,
    scale only axis,
    xlabel={$z$},
    ylabel={$\varepsilon(z)$},
    grid=both,
    grid style={line width=0.1pt, draw=gray!20},
    xmin=-7, xmax=0,
    ymin=-0.008, ymax=0.10,
    ytick={0, 0.02, 0.04, 0.06, 0.08, 0.10},
    yticklabels={0, 0.02, 0.04, 0.06, 0.08, 0.10},
    tick label style={font=\tiny},
    label style={font=\scriptsize},
    samples=350,
    domain=-7:0,
    title={(b) Error by $z$},
    title style={at={(0.5,1.02)}},
    legend style={font=\tiny, at={(0.02,0.80)}, anchor=north west, draw=none, fill=none},
    legend cell align=left,
]
\addplot[fill={rgb,255:red,0;green,119;blue,187}, fill opacity=0.15, draw=none, forget plot] {ln(1 + 2^x)/ln(2) - (1 + x - floor(x)) * 2^(floor(x))} \closedcycle;
\addplot[color={rgb,255:red,0;green,119;blue,187}, thick, mark=none] {ln(1 + 2^x)/ln(2) - (1 + x - floor(x)) * 2^(floor(x))};
\addlegendentry{no LUT}
\addplot[color={rgb,255:red,0;green,153;blue,136}, thick, mark=none] table[x=z, y=err] {intlse_lut_error.dat};
\addlegendentry{with LUT}
\addplot[color=peak1, only marks, mark=*, mark size=1.8pt, forget plot] coordinates {(-1, {ln(1 + 2^(-1))/ln(2) - 2^(-1)})};
\addplot[color=peak2, only marks, mark=*, mark size=1.8pt, forget plot] coordinates {(-2, {ln(1 + 2^(-2))/ln(2) - 2^(-2)})};
\addplot[color=peak3, only marks, mark=*, mark size=1.8pt, forget plot] coordinates {(-3, {ln(1 + 2^(-3))/ln(2) - 2^(-3)})};
\addplot[color=peak4, only marks, mark=*, mark size=1.8pt, forget plot] coordinates {(-4, {ln(1 + 2^(-4))/ln(2) - 2^(-4)})};
\addplot[color=peak5, only marks, mark=*, mark size=1.8pt, forget plot] coordinates {(-5, {ln(1 + 2^(-5))/ln(2) - 2^(-5)})};
\node[anchor=north west, font=\tiny, color={rgb,255:red,0;green,68;blue,136}] at (axis cs:-6.9, 0.095) {$\int_{-\infty}^{0}\!\varepsilon\,dz \approx 0.212$};
\end{axis}
\end{tikzpicture}%
\hfill%
\begin{tikzpicture}[baseline=(current bounding box.north)]
\begin{axis}[
    width=0.24\textwidth,
    height=0.18\textwidth,
    scale only axis,
    xlabel={$\hat{f}(z)$},
    ylabel={$\varepsilon(z)$},
    grid=both,
    grid style={line width=0.1pt, draw=gray!20},
    xmin=0, xmax=1.05,
    ymin=-0.005, ymax=0.10,
    ytick={0, 0.02, 0.04, 0.06, 0.08, 0.10},
    yticklabels={0, 0.02, 0.04, 0.06, 0.08, 0.10},
    tick label style={font=\tiny},
    label style={font=\scriptsize},
    title={(c) Error by $\hat{f}(z)$},
    title style={at={(0.5,1.02)}},
]
\addplot[color={rgb,255:red,0;green,119;blue,187}, thick, mark=none, samples=600, domain=-8:-0.001]
    ({(1 + x - floor(x)) * 2^(floor(x))},
     {ln(1 + 2^x)/ln(2) - (1 + x - floor(x)) * 2^(floor(x))});
\addplot[fill={rgb,255:red,0;green,119;blue,187}, fill opacity=0.10, draw=none, forget plot, samples=600, domain=-8:-0.001]
    ({(1 + x - floor(x)) * 2^(floor(x))},
     {ln(1 + 2^x)/ln(2) - (1 + x - floor(x)) * 2^(floor(x))})
    -- (axis cs:0,0) -- cycle;
\addplot[color=peak1, only marks, mark=*, mark size=1.8pt, forget plot] coordinates {(0.5, 0.08496)};
\addplot[color=peak2, only marks, mark=*, mark size=1.8pt, forget plot] coordinates {(0.25, 0.07168)};
\addplot[color=peak3, only marks, mark=*, mark size=1.8pt, forget plot] coordinates {(0.125, 0.04439)};
\addplot[color=peak4, only marks, mark=*, mark size=1.8pt, forget plot] coordinates {(0.0625, 0.02468)};
\addplot[color=peak5, only marks, mark=*, mark size=1.8pt, forget plot] coordinates {(0.03125, 0.01305)};
\end{axis}
\end{tikzpicture}
\caption{%
\textbf{(a)}~Exact $f(z) = \log_2(1{+}2^z)$ and approximation $\hat{f}(z)$.
\textbf{(b)}~Per-step error $\varepsilon(z)$: colored markers show peak values at integer endpoints $z = -n$; shaded area equals $\int_{-\infty}^{0}\varepsilon\,dz \approx 0.212$.
As $z \to -\infty$, $\varepsilon(z) \to 0$.
\textbf{(c)}~Parametric plot of $(\hat{f}(z),\,\varepsilon(z))$ as $z$ sweeps from $-\infty$ to~$0$; each arc is one integer period, and matching colored markers identify the same $z = -n$ peaks as in~(b). %
}
\label{fig:error-analysis}
\end{figure}
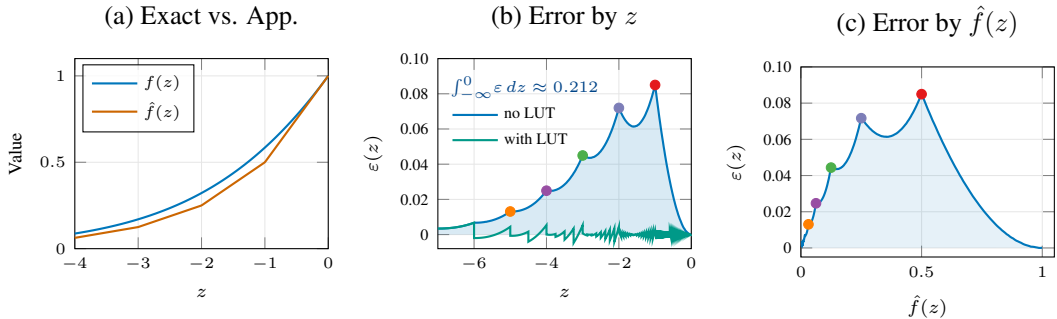
Each input float $x_i$ is converted to base-$2$ log space in Q16.16 fixed-point via $\mathrm{round}(x_i \cdot \log_2 e \cdot 2^{16})$. Pair-LSE then runs entirely in INT32: Eq.~\eqref{eq:combined-approx} is one add plus one shift, the LUT lookup is a single constant-memory load, and the warp- and block-level reductions are identical to \fuselse{}. Thread~$0$ converts the final integer back to float32 base-$e$ and writes it to global memory. The Q16.16 signed-integer representation bounds inputs to $|x|<32{,}768$ in log-base-2, which comfortably covers typical ML log-values.

\section{Individual kernel experiments and benchmarking}
\label{sec:experiments}
 
This section evaluates \fuselse{} and \intlse{} as standalone kernels in three versions: CUDA kernels for GPU, CPU implementations (multithreaded C++ for \fuselse{}, hand-vectorized AVX2 C for \intlse{}) that show when reducing arithmetic pays off on each platform, and reduced-precision CUDA variants.  \Cref{tab:kernels} lists all kernel variants and where they are used, and \cref{tab:robust-api} gives the supported shapes, dtypes, edge cases, backward pass, and fallback rules of the robust versions.

\subsection{Kernel benchmark}
\label{sec:exp-micro}
\textbf{Setup.}
We set $B = 512$, $N$ swept on a log grid from $2$ to $128{,}256$,
inputs drawn from $\mathrm{U}(-1, 0)$, $\mathrm{U}(-100, 0)$, and
$\mathrm{U}(-500, 0)$. We compare \texttt{torch.logsumexp},
\fuselse{}, and \intlse{} on GPU: V100-PCIe-32\,GB ($200$
iters, $20$ warm-ups) and CPU: Xeon Gold 6134 ($50/30/20/10$ iters scaled with $N$, $5$
warm-ups). Each point is the median over $20$ random seeds.
Error is reported against the FP64 baseline, with a
no-LUT \intlse{} ablation in the error panels.

\begin{figure}[!htbp]
  \centering
  \includegraphics[width=\linewidth]{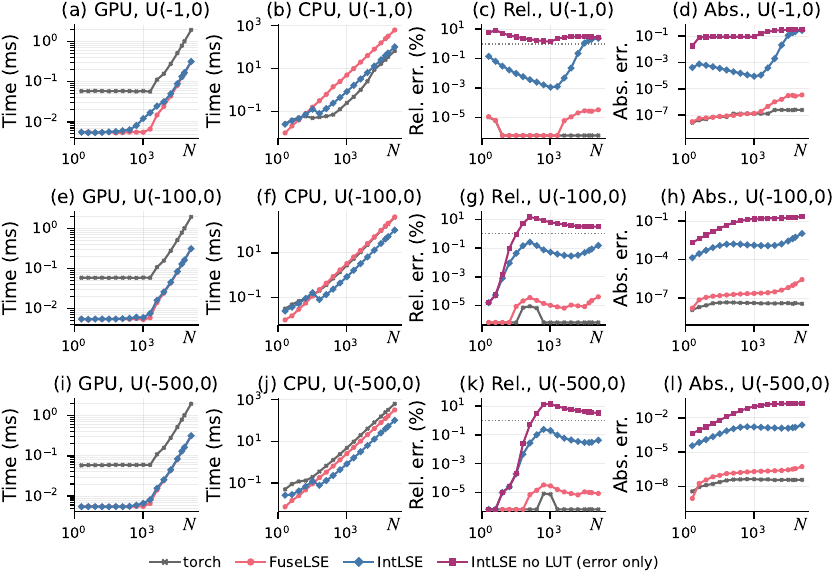}
  \caption{%
    Batched LSE microbenchmark, $B = 512$, $N$ swept $2$ to
    $128{,}256$. Rows: input ranges $\mathrm{U}(-1, 0)$,
    $\mathrm{U}(-100, 0)$, $\mathrm{U}(-500, 0)$. Columns: GPU
    per-call time (V100), CPU per-call time (Xeon Gold 6134), median relative error (\%), and mean
    absolute error vs.\ an FP64 inline reference; the no-LUT
    \intlse{} ablation is shown only in the error columns.}
  \label{fig:microbenchmark}
\end{figure}

\textbf{Results.}
On GPU (Fig.~\ref{fig:microbenchmark}, col.~1) \fuselse{} and
\intlse{} speed are within $1\%$ of each other and beat
\texttt{torch.logsumexp} by ${\sim}10\!-\!11\times$ for
$N \le 1024$ and ${\sim}6\times$ for $N \ge 8192$, settling at
$6.24\times/6.18\times$ at $N = 128{,}256$ ($1.94$\,ms vs.\
$0.310/0.313$\,ms).
On CPU (col.~2) \intlse{} speed beats \fuselse{} by
$3.1\!-\!6.1\times$; against torch the gap is range-dependent:
$1.54\times$ slower on $\mathrm{U}(-1,0)$, $3.66\times$ faster
on $\mathrm{U}(-100,0)$, and $6.15\times$ faster on
$\mathrm{U}(-500,0)$ at $N = 128{,}256$.
\fuselse{} matches torch at float precision
(median rel.\ err.\ ${<}10^{-4}\%$). \intlse{} with the LUT
stays below $0.2\%$ on the wider ranges ($0.15\%$ and $0.04\%$
at $N = 128{,}256$ for $\mathrm{U}(-100,0)$ and
$\mathrm{U}(-500,0)$) and grows to $2.38\%$ on
$\mathrm{U}(-1, 0)$; the no-LUT ablation saturates around
${\sim}0.2$ in absolute error at large $N$, and ${\sim}20\!-\!80\times$ above
the LUT-corrected variant on the wider ranges.

\textbf{Discussion.}
The two GPU regimes match the profile in \cref{sec:profile}: ${\sim}10\!-\!11\times$ at small $N$
comes from collapsing nine kernel launches into one, and ${\sim}6\times$ at large $N$ comes from cutting torch's HBM traffic from $6\times$ the
input volume to $1\times$. \fuselse{} and
\intlse{} coincide on GPUs because in both regimes the arithmetic cost of the pair function is hidden, so replacing \texttt{exp}/\texttt{log} with integer ops has no effect on
runtime. 
On CPU, there is no SFU and the kernel becomes compute-bound, so \texttt{exp}/\texttt{log} loses to \intlse{}'s integer arithmetic. 
The range-dependent gap against torch reflects how torch's Single Instruction Multiple Data (SIMD) pipeline responds to different input values, whereas the runtime of \intlse{}'s integer arithmetic is independent of the input values. On all but the narrowest range, \intlse{} is faster than torch even though it uses a single thread against torch's four. 

\Cref{tab:dist-accuracy} extends this comparison to additional input distributions, including extreme values: \fuselse{}'s p95 error stays within $4.5{\times}10^{-5}$, the same order as torch's own FP32 error, and \intlse{}'s within $4.2{\times}10^{-3}$. With matched single threads (\cref{tab:dist-cpu}), \intlse{} runs in a constant $27.6$\,ms on every finite input, $1.2\!-\!1.3\times$ faster than torch when no input underflows and $10.7\!-\!22.0\times$ faster otherwise.

\subsection{Cross-GPU time decomposition}
\label{sec:exp-cross-gpu-breakdown}
\begin{figure}[!htbp]
\centering
\includegraphics[width=\linewidth]{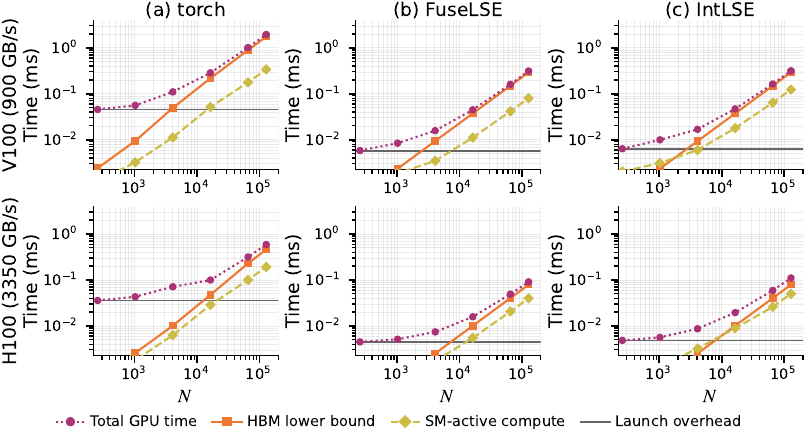}
\caption{Per-call time decomposition of \texttt{torch.logsumexp},
\fuselse{}, and \intlse{} (columns (a)/(b)/(c)) on V100 (top)
and H100 (bottom). All four
lines come from a single ncu run: total GPU time, HBM lower
bound, SM-active compute, and a launch reference (total time
at the smallest $N = 256$). Total time approaches the HBM
lower bound at large $N$, while SM-compute stays below it
across all kernels.}
\label{fig:cross-gpu-breakdown}
\end{figure}

\textbf{Setup.}
To verify the regime analysis of \cref{sec:exp-micro} and check how the speedup generalizes across GPUs, we decompose where each kernel's GPU time goes on V100-PCIe-32\,GB  ($900$\,GB/s peak HBM) and H100-SXM5-80\,GB ($3350$\,GB/s) using NVIDIA Nsight Compute. 
We profile the three kernels at $B = 512$, across $N \in \{256, 1024, 4096, 16384, 65536, 128256\}$, with $10$ launches per kernel, and report (Fig.~\ref{fig:cross-gpu-breakdown}) total GPU time, HBM lower bound (\texttt{dram\_\_bytes.sum} / peak HBM bandwidth), Streaming Multiprocessor (SM)-active compute time ($\sum_k \mathrm{gpu\_time}_k \cdot \mathrm{sm\_pct}_k$), and a launch reference (total GPU time at the smallest profiled $N = 256$). Inputs use $\mathrm{U}(-100, 0)$.

\textbf{Results.}
At small $N$, the total GPU time of all three kernels sits well
above both the HBM lower bound and the SM-compute line,
tracking the launch reference. At large $N$ it converges to
each kernel's HBM lower bound on both V100 and H100, with
SM-compute staying well below; \texttt{torch.logsumexp} moves
${\sim}6\times$ more bytes than both kernels.
\intlse{}'s integer pair costs ${\sim}50\%$ more SM-compute
than \fuselse{} on V100 ($123$ vs.\ $80\,\mu$s at
$N = 128{,}256$) and ${\sim}26\%$ more on H100 ($50$ vs.\
$40\,\mu$s).

\textbf{Discussion.}
For small $N$, the gap to the launch reference is due to kernel fusion; for large $N$, the ${\sim}6\times$ speedup comes from the amount of bytes moved, since
all three kernels saturate HBM well above their SM-compute
lines. \intlse{}'s extra SM-compute over \fuselse{} reflects the
absence of an integer advantage on GPUs.
\citet{jia2018dissecting} report V100 INT32 and FP32 share latency of
$4$-cycle. Even on older Pascal GPUs, SFU
instructions only (\texttt{\_\_expf}, \texttt{\_\_logf}) sit at
${\sim}14$ cycles, while \intlse{} adds ALU
cost on top. 
On CPUs, FP32 ADD latency is around 4× that of INT32 ADD of 1 cycle~\cite{agner2014instruction}, single \texttt{log} can cost up to $40$ cycles~\cite{shibata2019sleef}.
\fuselse{} is therefore the right choice on
modern GPUs; \intlse{}'s value is restricted to CPU and
platforms without a fast SFU.

We further compare against compiler-fused baselines, PyTorch~2.8 \texttt{torch.compile} and JAX~0.7.1 (XLA), on V100, H100, H200, and the newest B300, excluding compilation time (\cref{tab:compiler-baselines}). The speedup persists on newer GPUs: \fuselse{} is $6.3\!-\!6.9\times$ faster than eager PyTorch on all four GPUs and $1.8\!-\!2.1\times$ faster than both compilers; on B300, Triton and XLA cannot yet target \texttt{sm\_103}.
Nsight Compute explains the gap (\cref{tab:memory-profile}). Both compilers keep the two-pass structure and move $1.9\!-\!2.0\times$ the input volume per call, versus ${\sim}6\times$ for eager PyTorch and $1.01\times$ for \fuselse{}. The extra \texttt{log} calls do not offset this saving: the pair-wise \texttt{exp}/\texttt{log} work and the input traffic both scale as $O(N)$, so \fuselse{}'s arithmetic intensity is constant in $N$, and although its SFU utilization rises from $28\%$ (V100) to $46\%$ (H100) and $66\%$ (H200) as HBM bandwidth grows, at most $0.21\%$ of cycles stall waiting for a math pipe.

\subsection{Reduced-bit LSE}
\label{sec:16b}
As seen from \cref{sec:exp-micro} and \ref{sec:exp-cross-gpu-breakdown}, we address the launch overhead by using a single kernel launch in \fuselse{}, yet the HBM transfer bottleneck remains. To address this, we reduce the input bit-width and halve the bytes that must cross HBM. We evaluate two $16$-bit variants alongside the $32$-bit baselines: FuseLSE16 (FP16 storage, FP32
internal accumulation) and IntLSE-Q10.6 (INT16 Q10.6
storage, INT32 internal accumulation; with $10$ integer and $6$ fractional bits).

\begin{figure}[!htbp]
\centering
\includegraphics[width=\linewidth]{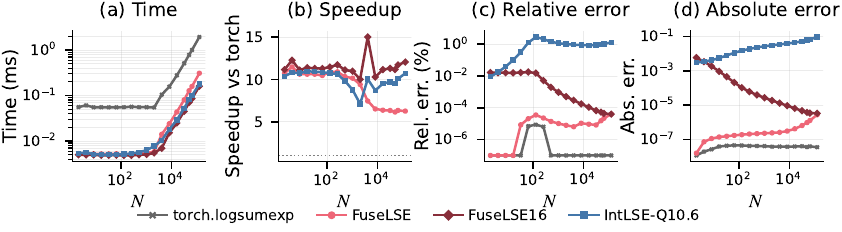}
\caption{%
  Reduced-bit \texttt{logsumexp} on V100-PCIe ($B\!=\!512$,
  $x\!\sim \mathrm{U}(-100,0)$).
  \textbf{(a)} Per-call time.
  \textbf{(b)} Speedup of the three fused variants vs.\
  \texttt{torch.logsumexp}.
  \textbf{(c)} Median relative error against FP64.
  \textbf{(d)} Mean absolute error against FP64.}
\label{fig:reduced-bit}
\end{figure}

\textbf{Setup.}
We set $B\!=\!512$, $N$ from $2$ to $128{,}256$,
inputs $\mathrm{U}(-100, 0)$. We compare the time and error of
\texttt{torch.logsumexp}, \fuselse{}, FuseLSE16, and
IntLSE-Q10.6 on V100-PCIe-32\,GB, Times are from CUDA events averaged over $1000$ iterations
after $100$ warm-ups, accuracy is averaged over $20$ seeds against FP64 reference, as seen in \cref{fig:reduced-bit}.

\textbf{Results.}
For small $N$ ($\le\!1024$) all three kernels are launch-overhead-bound and tie at $\sim\!10\!-\!11\times$. At $N\!=\!128{,}256$ the three kernels reach $6.25\times$ (\fuselse{}), $12.06\times$ (FuseLSE16), and $10.73\times$
(IntLSE-Q10.6) over \texttt{torch.logsumexp}. The two $16$-bit variants split on accuracy.
FuseLSE16's mean absolute error stays below
$3\!\times\!10^{-3}$ and decreases with $N$, tracking \texttt{torch.logsumexp} above
$N\!\approx\!10^4$. IntLSE-Q10.6's error instead grows
with $N$, reaching ${\sim}1\%$ at
$N = 128{,}256$, four orders of magnitude above
FuseLSE16.

\textbf{Discussion.}
On this V100 workload, FuseLSE16 dominates: same HBM saving, near FP32 accuracy.
IntLSE-Q10.6 captures the bandwidth speedup but is preferable
only on hardware without a fast SFU path, where the
integer pipeline becomes the actual compute floor. We do not pursue
either variant in the downstream applications, since
surrounding operations run in FP32.

\section{Applications}
\label{sec:applications}
We benchmark our kernels in three applications against the baselines used in
practice. RLHF log-probability computation (\cref{sec:exp-rlhf}, GPU) replaces \texttt{torch.logsumexp}
on dense $(B, V)$ vocabulary reductions. 
Probabilistic-circuit
inference (\cref{sec:exp-pyjuice}, GPU) replaces PyJuice's default and sparse-Triton sum-layer kernels with sparse variants of \fuselse{}/\intlse{}. In both RLHF and Probabilistic-circuit, \fuselse{} is our recommended kernel on GPU.
Particle filtering (\cref{sec:exp-pf}, CPU) is commonly deployed on CPUs, where \texttt{exp}/\texttt{log} are expensive without an SFU. We replace the NumPy LSE with the CPU version of \intlse{}.
\subsection{RLHF log-probability computation on GPU}
\label{sec:exp-rlhf}
Reinforcement-learning~\cite{lambert2025reinforcement} of LLMs computes per-token log-probabilities $z_{y_t} - \lse(z_1, \dots, z_V)$ over a vocabulary of size $V$, so each token costs one $V$-element LSE. The HuggingFace TRL library~\cite{vonwerra2020trl} packages this as \texttt{selective\_log\_softmax} (SLS), the operation we benchmark here.

\textbf{Setup.}
We benchmark across four vocabulary sizes spanning the modern open-model landscape: Llama-2~\cite{touvron2023llama} ($V\!=\!32\text{K}$), GPT-2~\cite{radford2019language} ($V\!=\!50\text{K}$), Llama-3~\cite{grattafiori2024llama} ($V\!=\!128\text{K}$), and Gemma~\cite{team2024gemma} ($V\!=\!256\text{K}$). Since SLS's cost depends only on the batch size, sequence length, and vocabulary size, not on the backbone's parameter count, we use a single GPT-2 (124\,M) backbone for the forward pass and swap only the LM head to match each target vocabulary size. For $V\!=\!50\text{K}$ we keep the pretrained head; for the other sizes we use a random projection of matching shapes.
The GPT-2 backbone runs a forward pass on Anthropic HH-RLHF text~\cite{bai2022training} tokenized to fixed length
$T\!=\!1024$, producing logits of shape $(B, T{-}1, V)$; we then time TRL's \texttt{selective\_log\_softmax}, which reduces
over the vocabulary axis $V$ at every position.
\fuselse{} and \intlse{} are plugged into TRL by reassigning \texttt{torch.logsumexp} at runtime to a wrapper that forwards
to our kernel and falls back to PyTorch otherwise; TRL itself is unmodified.
Batch sizes $B\!=\!(8, 8, 4, 2)$ scale with $V$ to fit GPU memory. We run $50$ timed iterations after $3$ warm-ups on V100-PCIe-32\,GB.
\vspace{-1em}
\begin{table}[!htbp]
\centering
\caption{Per-call timings of the GPT-2 LM-head + SLS stage (V100, $B{=}8$, $T{=}1024$, $V{=}50K$).}
\label{tab:rlhf-cmp}
\small
\setlength{\tabcolsep}{2pt}
\begin{tabular*}{\linewidth}{@{\extracolsep{\fill}}l cccccc@{}}
\toprule
                       & \texttt{torch} & \texttt{torch.compile} & \fuselse{}     & \intlse{}      & \texttt{cce} & \texttt{cce} (\texttt{torch.compile}) \\
\midrule
LM-head + SLS (ms)     & 59.12          & 51.64                  & 49.20          & 49.33          & 133.11       & \textbf{48.74} \\
SLS (ms)               & 12.34          & \emph{fused}           & \textbf{2.01}  & 2.07           & \emph{fused} & \emph{fused} \\
Cold compile (s)       & --             & 4.56                   & --             & --             & 0.97         & 1.74 \\
\bottomrule
\end{tabular*}
\end{table}

\begin{figure}[!htbp]
  \centering
  \includegraphics[width=\linewidth]{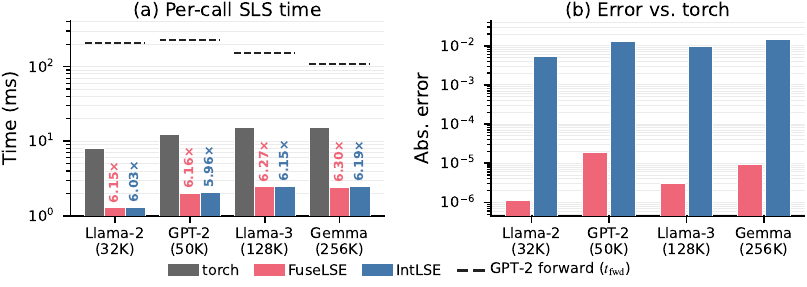}
  \caption{%
    RLHF SLS benchmark on V100-PCIe, GPT-2 backbone,
    $T\!=\!1024$, batch sizes $B\!=\!(8,8,4,2)$.
    \textbf{(a)} Per-call SLS time (log-scale $y$-axis): each
    call processes the full $(B, T{-}1, V)$ logits tensor and
    produces $B\!\cdot\!(T{-}1)$ per-token log-probabilities. Speedup over
    \texttt{torch.logsumexp} annotated above each \fuselse{} /
    \intlse{} bar in the matching colour; the dashed reference
    is the full GPT-2 forward time at each $V$.
    \textbf{(b)} Absolute reduction error vs.\
    \texttt{torch.logsumexp}.
    }
  \label{fig:rlhf-e2e}
\end{figure}

\textbf{Results.}
Both \fuselse{} and \intlse{} achieve $5.96$--$6.30\times$
speedup over \texttt{torch.logsumexp} across all four
vocabularies (\cref{fig:rlhf-e2e}, panel a), with the two
kernels within ${\sim}3.3\%$ of each other in every column.
The dashed reference is the GPT-2 forward time at the same
shape; SLS contributes to a significant fraction of time (${\sim}4\%$--$14\%$) and grows with $V$, so the kernel speedup gets more important as open-model vocabularies expand.
On accuracy (\cref{fig:rlhf-e2e}, panel b), \fuselse{} matches
\texttt{torch.logsumexp} to $1.1\!\times\!10^{-6}$--$1.9\!\times\!10^{-5}$
absolute error in the logsumexp output; \intlse{} stays within
$5.4\!\times\!10^{-3}$--$1.5\!\times\!10^{-2}$.
RLHF is the only setting in this paper where prior matmul-LSE-fused methods~\cite{wijmans2024cut} apply. On the GPT-2 LM-head + SLS benchmark (\cref{tab:rlhf-cmp}), \fuselse{} and \intlse{} deliver a $\sim\!6\times$ speedup on the SLS step and $1.20\times$ on the LM-head + SLS stage, capped by the shared 47-ms matmul. They pay no recompile cost, in contrast to \texttt{torch.compile} and \texttt{cce(torch.compile)} (\texttt{torch.compile} based implementation), which require $4.56/1.74\,$s of cold compile on every new input shape.

\subsection{Probabilistic circuit inference on GPU}
\label{sec:exp-pyjuice}
We evaluate \fuselse{} and \intlse{} for probabilistic circuits, where each sum layer computes $\lse$ over weighted child log-likelihoods.

\begin{figure}[!htbp]
  \centering
  \includegraphics[width=\linewidth]{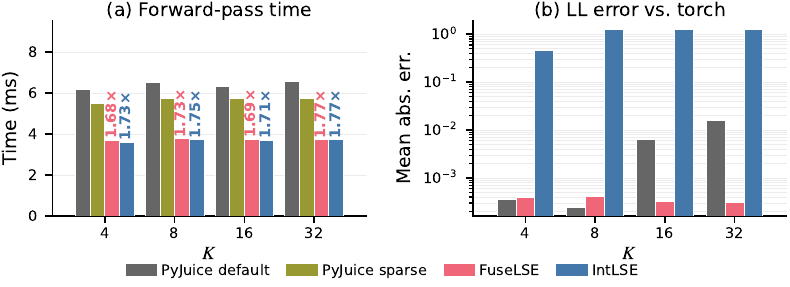}
  \caption{%
    PyJuice HCLT inference on MNIST, batch size $512$, V100-PCIe-32\,GB.
    \textbf{(a)} Forward-pass time per batch vs.\ number
    of latents $K$ for four implementations: PyJuice default
    (mixed-mode), PyJuice sparse (forced sparse),
    \fuselse{}, and \intlse{}. Speedup over PyJuice default is
    annotated above each \fuselse{} / \intlse{} bar in
    matching colour.
    \textbf{(b)} Per-sample mean absolute log-likelihood error
    against a \texttt{torch.logsumexp} reference (log
    axis), computed on the full MNIST validation set.}
  \label{fig:pyjuice}
\end{figure}

\textbf{Setup.}
We build Hidden Chow--Liu Tree (HCLT) circuits on MNIST
($28\!\times\!28$ pixels, $784$ features) with
PyJuice~\cite{liu2024scaling}. We sweep the latent-state count
$K\!\in\!\{4, 8, 16, 32\}$, producing circuits of increasing
width. In PyJuice's sparse mode, each sum layer dispatches a chain of torch operations that
performs a two-pass max-shift logsumexp ($\max$, then $\exp, \sum, \log$). We replace this with our fused kernel by reusing the \intlse{} and \fuselse{} pair primitives, but bundle a sparse gather and log-weight add inside one kernel pass. All layers are forced into sparse mode
for our kernels, so the replacement covers the full circuit. Forward-pass time is averaged over $20$ batches at batch size $512$ after a $5$-batch warm-up on Tesla V100-PCIe-32\,GB. For accuracy, we report the per-sample mean absolute log-likelihood error against a \texttt{torch.logsumexp} reference,
computed on the full MNIST validation set ($n = 9{,}728$ samples).

\textbf{Results.}
For throughput, \fuselse{} and \intlse{} achieve a $\sim1.7\times$ forward-pass speedup over the PyJuice default mode for all four $K$ (\cref{fig:pyjuice}, panel a), with the two kernels within ${\sim}3\%$ of each other in every column. Separate LSE speedup is invisible since the call is integrated sparsely inside the forward pass. 
For accuracy (\cref{fig:pyjuice}, panel b), \fuselse{} stays within $3.0$--$4.2 \!\times\! 10^{-4}$ absolute log-likelihood error, while \intlse{} sits at $0.45$--$1.26$ nats, under the per-sample LL of $\approx\!-4500$ nats. PyJuice's default LL relative to torch is shown for context, comparable to our \fuselse{} at small $K$ but growing to $1.6 \!\times\! 10^{-2}$ at $K\!=\!32$.

\subsection{Particle filter on CPU}
\label{sec:exp-pf}
We evaluate \intlse{} for the particle filter~\cite{chopin2020introduction, code_particles} on CPU, where particle filters are commonly deployed.
Each time step issues one length $N$
LSE call to normalise particle weights, so an $N$-particle,
$T$-step pass executes $T$ such calls.

\textbf{Setup.}
We use the univariate stochastic-volatility
model~\cite{kim1998stochastic} (default parameters
from~\cite{pitt1999filtering}) as packaged
by~\cite{chopin2020introduction}, with $T = 200$ observations. We sweep $N\!\in\!\{50, 100, 200, 500, 1000, 2000\}$ on Intel
Xeon Gold 6134, compared across: \texttt{scipy.special.logsumexp}, the NumPy (actual LSE path in~\cite{chopin2020introduction}),
\texttt{torch.logsumexp}, and
\intlse{}.
For accuracy, we report the mean absolute error against the FP64 LSE baseline, averaged
over $T$.
For timing, we report the per-call median wall-clock time, $100$ warm-ups followed by $500$ timed
calls.

\begin{figure}[!htbp]
  \centering
  \includegraphics[width=\linewidth]{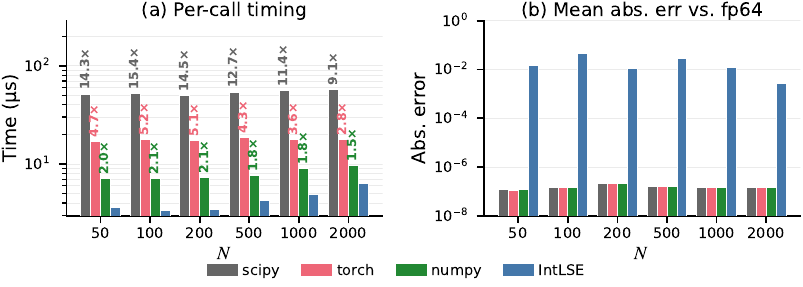}
\caption{%
    Particle-filter LSE primitive for the stochastic-volatility model of~\cite{pitt1999filtering} with $T{=}200$ observations, evaluated on Intel Xeon Gold 6134 with one thread for all methods. Inputs are the FP32 log weights encountered along the filter trajectory.
    \textbf{(a)} Median latency of one LSE call versus the number of particles $N$; labels report the speedup of \intlse{} over each baseline.
    \textbf{(b)} Mean absolute LSE error of SciPy, PyTorch, NumPy, and \intlse{} relative to an FP64 reference, averaged over all $T$ calls.
    The figure reports primitive-level latency and accuracy, not complete-filter runtime.%
    }
  \label{fig:pf}
\end{figure}
\textbf{Results.}
\intlse{} is the fastest primitive at every $N$
(\cref{fig:pf}, panel a): $9.1$--$15.4\times$ faster than
scipy, $1.5$--$2.1\times$ faster than
NumPy, and $2.8$--$5.2\times$ faster than
torch. Its absolute error stays ${\sim}10^{-2}$ across the $N$, making \intlse{} a viable LSE primitive for CPU particle
filters.

\section{Conclusion and discussion}
\label{sec:conclusion}
We presented \fuselse{} and \intlse{}, two kernels that serve as replacements for \texttt{torch.logsumexp}, delivering ${\sim}6\!-\!12\times$ speedups across microbenchmarks and three ML applications. Our experiments show that for memory-bound reductions like LSE and softmax on GPU, kernel fusion is the critical lever, cutting redundant HBM traffic that arithmetic optimization cannot reach. On platforms without a fast SFU (CPUs, microcontrollers, FPGAs), computation becomes the bottleneck and integer arithmetic wins.

\paragraph{Broader impact.}
For ML systems, redundant HBM traffic is the dominant cost of memory-bound reductions, and the analyses and kernels we present generalize to similar operations. For hardware designers, the lesson is that HBM bandwidth matters more than peak FLOPs for GPUs, while fast integer pipelines should still be exposed, not only on SFU-less platforms but also on GPUs. More fundamentally, modern GPUs are heavily optimized for dense matrix operations, a choice that suits today's LLMs but may not extend to future architectures.

\paragraph{Limitations.}
\intlse{}'s integer approximation incurs a cumulative error that may be unsuitable for the most accuracy-critical workloads. Our reduced-precision variants (FuseLSE16, IntLSE-Q10.6) capture additional bandwidth savings but are not yet integrated into the application benchmarks.

\begin{acknowledgements}
This work has been funded by the Research Council of Finland through the FerrARI project (Grant Number 359046). 
We acknowledge the computational resources provided by the Aalto Science-IT project.
This work was partially supported by the Wallenberg AI, Autonomous Systems and
Software Program (WASP) funded by the Knut and Alice Wallenberg Foundation.
\end{acknowledgements}
\newpage

\bibliographystyle{plainnat}
\bibliography{references}

\newpage
\appendix

\section{Technical appendices and supplementary material}
\subsection{\intlse{} Approximation-error analysis}
\label{app:error-details}

We split the expected accumulated error $\int_{-\infty}^{0}\!\varepsilon(z)\,dz$ in \cref{sec:intlse}, with $\varepsilon(z) = f(z) - \hat{f}(z)$, into the exact part (I), $f(z) = \log_2(1{+}2^z)$ and the approximation (II), $\hat{f}(z) = (1{+}F_z)\,2^{I_z}$:
\begin{equation}
\int_{-\infty}^{0}\!\varepsilon(z)\,dz = \underbrace{\int_{-\infty}^{0}\!\log_2(1{+}2^z)\,dz}_{(\mathrm{I})}
\;-\;
\underbrace{\int_{-\infty}^{0}\!(1{+}F_z)\,2^{I_z}\,dz}_{(\mathrm{II})}.
\end{equation}
For (II), summing over periods with $t = z + n \in [0,1)$,
\begin{equation}
    (\mathrm{II}) = \sum_{n=1}^{\infty} 2^{-n}\!\!\int_{0}^{1}\!(1{+}t)\,dt = \tfrac{3}{2}\sum_{n=1}^{\infty} 2^{-n} = \tfrac{3}{2}.
\end{equation}
For (I), substituting $u = 2^z$ and using
$\int_{0}^{1}\!\ln(1{+}u)/u\,du = \pi^2/12$,
\[
(\mathrm{I}) = \frac{1}{(\ln 2)^2}\!\!\int_{0}^{1}\!\frac{\ln(1{+}u)}{u}\,du
= \frac{\pi^2}{12(\ln 2)^2}.
\]
Subtracting (II) from (I) yields
\begin{equation}
    \int_{-\infty}^{0}\!\varepsilon(z)\,dz
    \;=\; \frac{\pi^{2}}{12(\ln 2)^{2}} - \frac{3}{2}
    \;\approx\; 0.212.
\end{equation}

\subsection{Additional experimental results}

\begin{table}[h]
\centering
\caption{All kernels, organised by variant and arithmetic, with the sections where they are used.}
\label{tab:kernels}
\small
\begin{tabular*}{\linewidth}{@{\extracolsep{\fill}}l ll ll@{}}
\toprule
 & \multicolumn{2}{c}{\fuselse{} (float \texttt{exp}/\texttt{log})} & \multicolumn{2}{c}{\intlse{} (integer + LUT)} \\
\cmidrule(lr){2-3} \cmidrule(l){4-5}
Variant & Kernel & Used in & Kernel & Used in \\
\midrule
GPU, FP32            & \fuselse{}        & \S4.1--\S4.3, \S5.1 & \intlse{}         & \S4.1--\S4.2, \S5.1 \\
GPU, 16-bit          & \fuselse{}16      & \S4.3               & \intlse{}-Q10.6   & \S4.3 \\
GPU, sparse PC  & \fuselse{}-sparse & \S5.2               & \intlse{}-sparse  & \S5.2 \\
CPU                  & \fuselse{} (CPU)  & \S4.1               & \intlse{} (CPU)   & \S4.1, \S5.3 \\
\midrule
Robust API           & \fuselse{}-robust (GPU) & \cref{tab:robust-api} & \intlse{}-robust (CPU) & \cref{tab:robust-api} \\
\bottomrule
\end{tabular*}
\end{table}

\begin{table}[h]
\centering
\caption{Edge-case handling, backward support, and fallback rules of the robust kernels. copy: contiguous copy; $\rightarrow$ref: redirected to the reference implementation (PyTorch or NumPy); $\times$: unsupported. Overhead vs.\ the non-robust kernel at $N{=}128{,}256$ (GPU: V100, $B{=}512$; CPU: Xeon Gold 6148, one thread, $B{=}128$).}
\label{tab:robust-api}
\small
\setlength{\tabcolsep}{2pt}
\begin{tabular*}{\linewidth}{@{\extracolsep{\fill}}l cc cc c ccc c c@{}}
\toprule
 & \multicolumn{2}{c}{Axis} & \multicolumn{2}{c}{Layout} & Dtype & \multicolumn{3}{c}{Output for} & \multirow{2}{*}{Backward} & \multirow{2}{*}{Overhead} \\
\cmidrule(lr){2-3} \cmidrule(lr){4-5} \cmidrule(lr){6-6} \cmidrule(lr){7-9}
Kernel & Other & Multi & Non-contig. & $N{\le}1$ & Others & All $-\infty$ & $+\infty$ & NaN & & \\
\midrule
\fuselse{}-robust & $\rightarrow$ref & $\times$ & copy & $\rightarrow$ref & $\rightarrow$ref & $-\infty$ & $+\infty$        & NaN              & $\checkmark$           & +3.2\% \\
\intlse{}-robust  & $\rightarrow$ref & $\times$ & copy & $\rightarrow$ref & $\rightarrow$ref & $-\infty$ & $\rightarrow$ref & $\rightarrow$ref & $\rightarrow$ref & +7.4\% \\
\bottomrule
\end{tabular*}
\end{table}

\begin{table}[h]
\centering
\caption{Accuracy across input distributions: p95 absolute error vs.\ FP64 at $N{=}128{,}256$, $B{=}128$, 3 seeds. \fuselse{} on V100, \intlse{} on CPU; torch FP32 shown for reference. $^\dagger$Robust kernels.}
\label{tab:dist-accuracy}
\small
\begin{tabular*}{\linewidth}{@{\extracolsep{\fill}}l ccc@{}}
\toprule
Input & torch (FP32) & \fuselse{} & \intlse{} \\
\midrule
$\mathrm{U}(-1,0)$                   & 7.0e-7 & 4.1e-6 & 1.4e-4 \\
$\mathrm{U}(-100,0)$                 & 2.6e-7 & 3.0e-6 & 1.4e-3 \\
$\mathrm{U}(-500,0)$                 & 2.5e-7 & 8.7e-7 & 1.7e-3 \\
\midrule
$\mathcal{N}(0,1)$                   & 6.9e-7 & 8.4e-6 & 6.5e-4 \\
$\mathcal{N}(0,20^2)$                & 3.7e-6 & 7.7e-6 & 4.2e-3 \\
$\mathrm{U}(-1000,1000)$             & 2.9e-5 & 4.5e-5 & 2.1e-3 \\
Near ties, $\mathcal{N}(0,10^{-8})$  & 7.9e-7 & 4.1e-6 & 4.8e-4 \\
One $+1000$ entry                    & 0      & 0      & 6.1e-5 \\
50\% $-\infty$ mask$^\dagger$        & 7.0e-7 & 4.2e-6 & 3.0e-4 \\
\bottomrule
\end{tabular*}
\end{table}

\begin{table}[h]
\centering
\caption{CPU time across input distributions at $N{=}128{,}256$, $B{=}128$, on Xeon Gold 6148 with one thread for both methods. Underflow: share of $x_i - \max_j x_j < -87.3$, where FP32 \texttt{exp} is subnormal or zero. $^\dagger$Robust \intlse{} kernel.}
\label{tab:dist-cpu}
\small
\begin{tabular*}{\linewidth}{@{\extracolsep{\fill}}l cccc@{}}
\toprule
Input & Underflow (\%) & torch (ms) & \intlse{} (ms) & Speedup \\
\midrule
$\mathrm{U}(-1,0)$                   & 0.0   & 33.0  & 27.6 & 1.2$\times$  \\
$\mathrm{U}(-100,0)$                 & 12.7  & 294.4 & 27.6 & 10.7$\times$ \\
$\mathrm{U}(-500,0)$                 & 82.5  & 389.1 & 27.6 & 14.1$\times$ \\
\midrule
$\mathcal{N}(0,1)$                   & 0.0   & 34.2  & 27.6 & 1.2$\times$  \\
$\mathcal{N}(0,20^2)$                & 50.2  & 606.9 & 27.6 & 22.0$\times$ \\
$\mathrm{U}(-1000,1000)$             & 95.6  & 358.6 & 27.6 & 13.0$\times$ \\
Near ties, $\mathcal{N}(0,10^{-8})$  & 0.0   & 36.3  & 27.6 & 1.3$\times$  \\
One $+1000$ entry                    & 100.0 & 304.5 & 27.6 & 11.0$\times$ \\
50\% $-\infty$ mask$^\dagger$        & --    & 188.8 & 29.9 & 6.3$\times$  \\
\bottomrule
\end{tabular*}
\end{table}

\begin{table}[h]
\centering
\caption{LSE latency (ms) at $(B,N)=(512,128{,}256)$, excluding compilation; \fuselse{} speedup in parentheses. --: toolchain does not support B300.}
\label{tab:compiler-baselines}
\small
\begin{tabular*}{\linewidth}{@{\extracolsep{\fill}}l cccc@{}}
\toprule
Implementation & V100 & H100 & H200 & B300 \\
\midrule
Eager PyTorch          & 1.936 (6.25$\times$) & 0.584 (6.53$\times$) & 0.462 (6.86$\times$) & 0.366 (6.70$\times$) \\
\texttt{torch.compile} & 0.605 (1.95$\times$) & 0.163 (1.83$\times$) & 0.120 (1.78$\times$) & -- \\
JAX/XLA                & 0.651 (2.10$\times$) & 0.184 (2.06$\times$) & 0.134 (1.99$\times$) & -- \\
\fuselse{}             & \textbf{0.310}       & \textbf{0.0894}      & \textbf{0.0674}      & \textbf{0.0546} \\
\bottomrule
\end{tabular*}
\end{table}

\begin{table}[h]
\centering
\caption{Nsight Compute profile at $(B,N)=(512,128{,}256)$. Below each GPU: peak HBM bandwidth and estimated peak SFU throughput (SMs $\times$ 16 ops/clock $\times$ boost clock). Traffic: DRAM bytes per call relative to the input size. DRAM, SFU: achieved throughput as \% of peak. Pipe stall: \% of cycles in which warps wait for a busy math pipe (including the SFU). Multi-kernel values are time-weighted.}
\label{tab:memory-profile}
\small
\begin{tabular*}{\linewidth}{@{\extracolsep{\fill}}ll ccccc@{}}
\toprule
GPU & Implementation & Traffic & Time ($\mu$s) & DRAM (\%) & SFU (\%) & Pipe stall (\%) \\
\midrule
\multirow{4}{*}{\makecell[l]{V100\\\scriptsize 0.90 TB/s\\\scriptsize 1.8 Top/s SFU}}
                      & Eager PyTorch          & 5.98$\times$ & 1951.2 & 89.7 & 2.1  & 1.31 \\
                      & \texttt{torch.compile} & 1.98$\times$ & 618.8  & 93.5 & 7.8  & 2.18 \\
                      & JAX/XLA                & 2.02$\times$ & 656.0  & 90.0 & 6.4  & 6.06 \\
                      & \fuselse{}             & \textbf{1.01}$\times$ & \textbf{314.9} & 92.0 & 28.0 & 0.20 \\
\midrule
\multirow{4}{*}{\makecell[l]{H100\\\scriptsize 3.35 TB/s\\\scriptsize 4.2 Top/s SFU}}
                      & Eager PyTorch          & 5.88$\times$ & 594.4 & 77.5 & 3.3  & 0.64 \\
                      & \texttt{torch.compile} & 1.91$\times$ & 169.6 & 88.2 & 12.2 & 0.38 \\
                      & JAX/XLA                & 2.03$\times$ & 187.2 & 84.8 & 11.3 & 0.25 \\
                      & \fuselse{}             & \textbf{1.01}$\times$ & \textbf{94.2} & 84.0 & 45.8 & 0.10 \\
\midrule
\multirow{4}{*}{\makecell[l]{H200\\\scriptsize 4.80 TB/s\\\scriptsize 4.2 Top/s SFU}}
                      & Eager PyTorch          & 5.85$\times$ & 481.4 & 65.0 & 4.7  & 1.18 \\
                      & \texttt{torch.compile} & 1.89$\times$ & 124.6 & 81.0 & 19.1 & 1.14 \\
                      & JAX/XLA                & 2.03$\times$ & 141.0 & 77.1 & 16.7 & 0.57 \\
                      & \fuselse{}             & \textbf{1.01}$\times$ & \textbf{72.6} & 74.3 & 65.9 & 0.21 \\
\bottomrule
\end{tabular*}
\end{table}

\clearpage
\section*{NeurIPS Paper Checklist}

The checklist is designed to encourage best practices for responsible machine learning research, addressing issues of reproducibility, transparency, research ethics, and societal impact. Do not remove the checklist: {\bf The papers not including the checklist will be desk rejected.} The checklist should follow the references and follow the (optional) supplemental material.  The checklist does NOT count towards the page
limit. 

Please read the checklist guidelines carefully for information on how to answer these questions. For each question in the checklist:
\begin{itemize}
    \item You should answer \answerYes{}, \answerNo{}, or \answerNA{}.
    \item \answerNA{} means either that the question is Not Applicable for that particular paper or the relevant information is Not Available.
    \item Please provide a short (1--2 sentence) justification right after your answer (even for \answerNA). 
\end{itemize}

{\bf The checklist answers are an integral part of your paper submission.} They are visible to the reviewers, area chairs, senior area chairs, and ethics reviewers. You will also be asked to include it (after eventual revisions) with the final version of your paper, and its final version will be published with the paper.

The reviewers of your paper will be asked to use the checklist as one of the factors in their evaluation. While \answerYes{} is generally preferable to \answerNo{}, it is perfectly acceptable to answer \answerNo{} provided a proper justification is given (e.g., error bars are not reported because it would be too computationally expensive'' or ``we were unable to find the license for the dataset we used''). In general, answering \answerNo{} or \answerNA{} is not grounds for rejection. While the questions are phrased in a binary way, we acknowledge that the true answer is often more nuanced, so please just use your best judgment and write a justification to elaborate. All supporting evidence can appear either in the main paper or the supplemental material, provided in appendix. If you answer \answerYes{} to a question, in the justification please point to the section(s) where related material for the question can be found.

IMPORTANT, please:
\begin{itemize}
    \item {\bf Delete this instruction block, but keep the section heading ``NeurIPS Paper Checklist"},
    \item  {\bf Keep the checklist subsection headings, questions/answers and guidelines below.}
    \item {\bf Do not modify the questions and only use the provided macros for your answers}.
\end{itemize}


\begin{enumerate}

\item {\bf Claims}
    \item[] Question: Do the main claims made in the abstract and introduction accurately reflect the paper's contributions and scope?
    \item[] Answer: \answerYes{} 
    \item[] Justification: The two Kernels claimed in the abstract and introduction, their design can be seen in\cref{sec:method}, their speed up together with analysis in \cref{sec:experiments} and \cref{sec:applications}.

\item {\bf Limitations}
    \item[] Question: Does the paper discuss the limitations of the work performed by the authors?
    \item[] Answer: \answerYes{} 
    \item[] Justification: Limitations subsection can be found in \cref{sec:conclusion}.

\item {\bf Theory assumptions and proofs}
    \item[] Question: For each theoretical result, does the paper provide the full set of assumptions and a complete (and correct) proof?
    \item[] Answer: \answerYes{} 
    \item[] Justification: The theoretical proof of \intlse{} expected error is detailed in \cref{app:error-details}, with assumption in \cref{sec:intlse}.

    \item {\bf Experimental result reproducibility}
    \item[] Question: Does the paper fully disclose all the information needed to reproduce the main experimental results of the paper to the extent that it affects the main claims and/or conclusions of the paper (regardless of whether the code and data are provided or not)?
    \item[] Answer: \answerYes{} 
    \item[] Justification: All information needed to reproduce the main experimental results of the paper can be seen in \cref{sec:experiments} and \cref{sec:applications}.

\item {\bf Open access to data and code}
    \item[] Question: Does the paper provide open access to the data and code, with sufficient instructions to faithfully reproduce the main experimental results, as described in supplemental material?
    \item[] Answer: \answerYes{} 
    \item[] Justification: We included the code URL in this paper.

\item {\bf Experimental setting/details}
    \item[] Question: Does the paper specify all the training and test details (e.g., data splits, hyperparameters, how they were chosen, type of optimizer) necessary to understand the results?
    \item[] Answer: \answerYes{} 
    \item[] Justification: Our contributions are about computation speed up, not about training and testing, whose parameters are default from the original libraries; all details can be seen in \cref{sec:experiments} and \cref{sec:applications}.

\item {\bf Experiment statistical significance}
    \item[] Question: Does the paper report error bars suitably and correctly defined or other appropriate information about the statistical significance of the experiments?
    \item[] Answer: \answerYes{} 
    \item[] Justification: We report errors according to conventional methods; details can be seen in \cref{sec:experiments} and \cref{sec:applications}.

\item {\bf Experiments compute resources}
    \item[] Question: For each experiment, does the paper provide sufficient information on the computer resources (type of compute workers, memory, time of execution) needed to reproduce the experiments?
    \item[] Answer: \answerYes{} 
    \item[] Justification: We provide sufficient information on computer resources(CPU, GPU and memory) for each experiment, details can be seen in \cref{sec:experiments} and \cref{sec:applications}.
    
\item {\bf Code of ethics}
    \item[] Question: Does the research conducted in the paper conform, in every respect, with the NeurIPS Code of Ethics \url{https://neurips.cc/public/EthicsGuidelines}?
    \item[] Answer: \answerYes{} 
    \item[] Justification: The research conducted in this paper conform, in every respect, with the NeurIPS Code of Ethics \url{https://neurips.cc/public/EthicsGuidelines}.

\item {\bf Broader impacts}
    \item[] Question: Does the paper discuss both potential positive societal impacts and negative societal impacts of the work performed?
    \item[] Answer: \answerNA{} 
    \item[] Justification: This paper is not tied to particular applications or deployment; therefore, no societal impact.
    
\item {\bf Safeguards}
    \item[] Question: Does the paper describe safeguards that have been put in place for responsible release of data or models that have a high risk for misuse (e.g., pre-trained language models, image generators, or scraped datasets)?
    \item[] Answer: \answerNA{} 
    \item[] Justification: This paper does not introduce new data or models and poses no such risks.

\item {\bf Licenses for existing assets}
    \item[] Question: Are the creators or original owners of assets (e.g., code, data, models), used in the paper, properly credited and are the license and terms of use explicitly mentioned and properly respected?
    \item[] Answer: \answerYes{} 
    \item[] Justification: The original owners of the code this paper uses are cited.

\item {\bf New assets}
    \item[] Question: Are new assets introduced in the paper well documented and is the documentation provided alongside the assets?
    \item[] Answer: \answerYes{} 
    \item[] Justification: We included the code URL in this paper.

\item {\bf Crowdsourcing and research with human subjects}
    \item[] Question: For crowdsourcing experiments and research with human subjects, does the paper include the full text of instructions given to participants and screenshots, if applicable, as well as details about compensation (if any)? 
    \item[] Answer: \answerNA{} 
    \item[] Justification: This paper does not involve crowdsourcing nor research with human subjects.

\item {\bf Institutional review board (IRB) approvals or equivalent for research with human subjects}
    \item[] Question: Does the paper describe potential risks incurred by study participants, whether such risks were disclosed to the subjects, and whether Institutional Review Board (IRB) approvals (or an equivalent approval/review based on the requirements of your country or institution) were obtained?
    \item[] Answer: \answerNA{} 
    \item[] Justification: This paper does not involve crowdsourcing nor research with human subjects.

\item {\bf Declaration of LLM usage}
    \item[] Question: Does the paper describe the usage of LLMs if it is an important, original, or non-standard component of the core methods in this research? Note that if the LLM is used only for writing, editing, or formatting purposes and does \emph{not} impact the core methodology, scientific rigor, or originality of the research, declaration is not required.
    \item[] Answer: \answerNA{} 
    \item[] Justification: In this paper, LLM is used only for writing, and formatting purposes and does not impact the core methodology, scientific rigor, or originality of the research.

\end{enumerate}

\end{document}